\documentclass[prb,twocolumn,showpacs]{revtex4}
\usepackage{amsfonts}

\begin{document}

\title  {Replica field theories, Painlev\'e transcendents, and
        exact correlation functions}

\author {Eugene Kanzieper}
\email  {eugene.kanzieper@weizmann.ac.il}

\affiliation
        {Department of Condensed Matter Physics,
        Weizmann Institute of Science, Rehovot 76100, Israel}
\date   {July 31, 2002}

\begin  {abstract}

Exact solvability is claimed for nonlinear replica $\sigma$ models
derived in the context of random matrix theories. Contrary to
other approaches reported in the literature, the framework
outlined does not rely on traditional ``replica symmetry
breaking'' but rests on a previously unnoticed exact relation
between replica partition functions and Painlev\'e transcendents.
While expected to be applicable to matrix models of arbitrary
symmetries, the method is used to treat fermionic replicas for the
Gaussian unitary ensemble (GUE), chiral GUE (symmetry classes
{\texttt A} and {\texttt AIII} in Cartan classification) and
Ginibre's ensemble of complex non-Hermitean random matrices.
Further applications are briefly discussed.

\end{abstract}

 \maketitle

Replica field theories are notoriously known for subtleties
involved in performing the replica limit, $n \rightarrow 0$,
devised \cite{EA-1975} to carve a physical observable of interest
out of the ``annealed'' average
\begin  {eqnarray}
        \label{quenched}
        Z_n(\epsilon_1,\cdots,\epsilon_p) =
        \left<
        \prod_{k=1}^p
        \det{}^n (\epsilon_k - {\bf {\cal H}}) \right>
\end    {eqnarray}
often called the replica partition function; generically, $ {\rm
Im} \, \epsilon_k \neq 0 $. The average $\langle \cdots \rangle$
runs over an ensemble of stochastic Hamiltonians ${\bf \cal H}$
which, throughout this Letter, will be modeled by random matrices
\cite{M-1991} of prescribed symmetries. Given (\ref{quenched}),
spectral properties of ${\bf {\cal H}}$ can be obtained from the
$p$-point Green's function $G(\epsilon_1,\cdots,\epsilon_p)=
\langle \prod_{k=1}^p {\rm tr} \, (\epsilon_k - {\cal H})^{-1}
\rangle$ for which the replica limit reads
\begin  {eqnarray}
        \label{p-cf}
        G(\epsilon_1,\cdots,\epsilon_p) =
        \lim_{n\rightarrow 0}
        \frac{1}{n^p}
        \frac{\partial^p}{\partial
        \epsilon_1\cdots\partial\epsilon_p}
        Z_n(\epsilon_1,\cdots,\epsilon_p).
\end    {eqnarray}
Equation (\ref{p-cf}) assumes mutual commutativity of the
following operations: disorder averaging, differentiation, the
replica limit and a thermodynamic limit, if necessary. Being an
undoubtedly correct mathematical identity under the conditions of
commutativity, the recipe (\ref{p-cf}) may become a pitfall
\cite{HP-1979,VZ-1985} if applied unconsciously in the context of
replica field theories.

A difficulty lurks behind a field theoretic prescription to
compute the partition function (\ref{quenched}). Sketchy, an
original random system is substituted by its $|n|$ identical
noninteracting copies, or replicas. Each copy, exemplified by a
single determinant $\det(\epsilon - {\bf \cal H})$, is represented
by a functional integral over an auxiliary field which is either
bosonic or fermionic by nature depending on the sign of $n$.
Exponentiating a random Hamiltonian ${\bf \cal H}$, such a
representation facilitates a nonperturbative averaging over an
ensemble of stochastic Hamiltonians in (\ref{quenched}) and
eventually results in effective field theories defined on either a
compact \cite{ELK-1980} ($n>0$) or a noncompact \cite{W-1979}
($n<0$) manifold.

The point to make is this: Since the number $|n|$ of functional
field integrals involved is a positive integer, $|n| \in {\mathbb
N}$, the validity of the resulting representation of the replica
partition function $Z_n$ is restricted to $|n| \in {\mathbb N}$,
too. Unfortunately, this is not enough to perform the replica
limit (\ref{p-cf}) determined by the behaviour of $Z_n$ in a close
vicinity of $n=0$. Therefore, a procedure of analytic continuation
of $Z_n$ away from $|n|$ positive integers is called for.

{\it Kamenev-M\'ezard's prescription}.---To keep discussion
concrete, let us reconsider a problem \cite{EJ-1976} of evaluation
of the one-point Green's function $G(\epsilon)$ for random
matrices with Gaussian distributed entries. In what follows, we
assume the Hamiltonian ${\cal H}$ to be drawn from the Gaussian
Unitary Ensemble \cite{M-1991} (${\rm GUE}_N$) specified by the
probability density $P_N({\bf \cal H}) \propto \exp(-{\rm tr}
\,{\bf \cal H}^2)$ for an $N \times N$ complex Hermitean matrix
${\cal H}$ to occur; $N \in {\mathbb N}$. The Green's function
$G(\epsilon)$ is determined by the replica limit
$G(\epsilon)=\lim_{n\rightarrow 0} n^{-1} \partial
Z_n(\epsilon)/\partial \epsilon$ with
\begin  {eqnarray}
        \label{GUE-Z}
        Z_n(\epsilon) =
        \left<
        \det{}^n (\epsilon - {\bf \cal H})
        \right>_{{\bf \cal H} \in {\rm GUE}_N}, \;\;
        n \in {\mathbb C}.
\end    {eqnarray}
Angular brackets $\langle \cdots \rangle$ stand for a matrix
integral over the properly normalised measure \cite{M-1991}
$P_N({\cal H}){\cal D H}$.

In the above definition of $Z_n$, the parameter $n$ is allowed to
be an arbitrary complex number ${\mathbb C}$. Therefore, if
applied directly to (\ref{GUE-Z}), the replica limit would result
\cite{K-2001} in a correct answer for $G(\epsilon)$. This is not
so, however, if one maps the partition function $Z_n$ onto a
variant \cite{KM-1999} of the replica $\sigma$ model
\cite{ELK-1980}, $Z_n \mapsto {\widetilde Z}_n$. Following the
standard steps of fermionic mapping, one derives
\cite{KM-1999,Remark1}
\begin  {eqnarray}
        \label{GUE-Z-mapped}
        {\widetilde Z}_n(\epsilon) = \left<
        \det{}^N (i\epsilon - {\bf \cal Q})
        \right>_{{\cal Q} \in {\rm GUE}_n}, \;\;
        n \in {\mathbb N}.
\end    {eqnarray}
Contrary to the starting point (\ref{GUE-Z}), the replica
parameter $n$ in (\ref{GUE-Z-mapped}) is now restricted to
positive integers {\it by derivation}. As a result, any attempt to
reconstruct the Green's function $G(\epsilon)$ out of ${\widetilde
Z}_n$ through the replica limit (\ref{p-cf}) will inevitably face
the problem of analytic continuation of ${\widetilde Z}_n$ away
from $n \in {\mathbb N}$.

Despite numerous efforts and discussions throughout more than two
decades, no mathematically satisfactory idea was brought in, even
though there exists a recipe \cite{KM-1999,KM-1999a} to deal with
the problem. Its detailed exposition can be found in Ref.
\cite{Z-1999}; below we only recollect the facts needed for
further discussion.

(i) As a prerequisite, one attempts to unveil an explicit
dependence of ${\widetilde Z}_n$ on the replica index $n$ which is
only implicit in (\ref{GUE-Z-mapped}). To this end, the matrix
integral (\ref{GUE-Z-mapped}) is evaluated {\it approximately}
through a saddle point procedure which makes sense if the
dimension $N$ of the random matrix ${\cal H}$ is large enough. For
not too large replica parameter $n \in {\mathbb N}$ (in
particular, $n$ should not scale with $N$), this yields
\cite{Z-1999}
        $
        {\widetilde Z}_n^{({\rm sp})}(\epsilon) \simeq \sum_{p=0}^n \texttt{vol}
        \left( G_{n,p} \right) \, z_{n,p}(\epsilon)
        $,
where
$\texttt{vol}
 \left(
 G_{n,p}
 \right)
 =
 \prod_{j=1}^p
 [\Gamma(j)/\Gamma(n-j+1)]
$ is the volume of Grassmannian  $G_{n,p} ={\rm U}(n)/{\rm U}(n-p)
\times {\rm U}(p)$, and $z_{n,p}(\epsilon)$ is a known function
(its explicit form is irrelevant to our discussion). The inner
index $p$ in the above sum $\sum_{p=0}^n(\cdots)$ counts a
hierarchy of (nonequivalent) causal \cite{Z-1999} saddle points
contributing the integral (\ref{GUE-Z-mapped}) over Hermitean
matrix ${\bf \cal Q}$; inequivalence of the saddle points
highlights a phenomenon of ``replica symmetry breaking''.

(ii) Further, one seeks a proper analytic continuation of
${\widetilde Z}_n^{({\rm sp})}$ away from $n \in {\mathbb N}$. The
(most successful so far) procedure devised in Ref. \cite{KM-1999}
suggests extending the summation over $p$ to infinity,
$\sum_{p=0}^n (\cdots) \mapsto \sum_{p=0}^\infty(\cdots)$, as the
group volumes $\texttt{vol}\left( G_{n,p} \right)$ vanish for $p
\ge n+1$. Such a proposal suffers from two major drawbacks
\cite{Z-1999}: (a) for $n \notin {\mathbb N}$ the group volumes
grow too fast with $p$ for the sum $\sum_{p=0}^\infty(\cdots)$ to
converge; (b) so extended to infinity the sum over $p$ would
necessarily involve a contribution from $n \sim {\cal O}(N)$ where
the summand is no longer given by $\texttt{vol} \left( G_{n,p}
\right) \, z_{n,p}(\epsilon)$. This questions the self-consistency
of the method as a whole making it somewhat deficient. Despite all
these drawbacks, however, the approach furnishes \cite{KM-1999} a
correct result for the large-$N$ GUE density of states in both
leading and subleading orders in $1/N$.

{\it Integrable hierarchies and exact Painlev\'e reduction}.---On
brief reflection, one has to admit that the {\it approximate}
evaluation of ${\widetilde Z}_n(\epsilon)$ is the key point to
blame for inconsistencies encountered in the procedure of analytic
continuation. For this reason, we opt a route based on {\it exact}
and, therefore, truly nonperturbative evaluation of the replica
partition function. As improbable or fantastic as it sounds, this
is not an impossible task.

Our claim of exact solvability of the replica model
(\ref{GUE-Z-mapped}) and the models of the same ilk rests on two
observations. (i) To make the first, we routinely reduce the
average over ${\cal Q} \in {\rm GUE}_n$ in (\ref{GUE-Z-mapped}) to
the $n$-fold integral \cite{KM-1999}
\begin  {eqnarray}
        \label{GUE-Z-nfold}
        {\widetilde Z}_n(\epsilon) = \int_{-\infty}^{+\infty}
        \prod_{k=1}^n d\lambda_k \, e^{-\lambda_k^2} (\lambda_k - i
        \epsilon)^N \Delta_n^2(\lambda).
\end    {eqnarray}
Here, $\Delta_n(\lambda)=\prod_{k>\ell=1}^n (\lambda_k -
\lambda_\ell)=\det(\lambda_k^{\ell-1})$ is the Vandermonde
determinant \cite{M-1991} which makes it possible to bring
(\ref{GUE-Z-nfold}) to the form \cite{K-2001}
\begin  {eqnarray}
        \label{n-factorisation}
        {\widetilde Z}_n(\epsilon) = e^{n \epsilon^2}
        {\widetilde \tau}_n(\epsilon; N), \;\;\; n \in {\mathbb
        N},
\end    {eqnarray}
which involves the H\"ankel determinant $
        {\widetilde \tau}_n(\epsilon; N) = \det
        \left[
        \partial^{k+\ell}_\epsilon\, {\widetilde \tau}_1(\epsilon; N)
        \right]_{k,\ell = 0, \cdots, n-1}$. The latter, as had first
been shown by Darboux \cite{Darboux} a century ago, satisfies the
equation \cite{Remark2}
\begin  {eqnarray}
        \label{Toda}
        {\widetilde \tau}_n \, {\widetilde \tau}^{\prime\prime}_n
        -
        ({\widetilde \tau}^\prime_n)^2 = {\widetilde \tau}_{n-1}
        {\widetilde \tau}_{n+1}, \;\;\; n \ge 1,
\end    {eqnarray}
given the initial conditions ${\widetilde \tau}_0 \equiv 1$ and
${\widetilde \tau}_1 = e^{-\epsilon^2}{\widetilde Z}_1(\epsilon)$;
the prime stands for $d/d\epsilon$. The structure of (\ref{Toda})
is eventually due to the $\beta=2$ symmetry of the replica field
theory encoded into $\Delta_n^2$ in (\ref{GUE-Z-nfold}); $\beta$
is Dyson's index. Equations (\ref{n-factorisation}) and
(\ref{Toda}) establish a hierarchy between nonperturbative replica
partition functions ${\widetilde Z}_n$ with different $n \in
{\mathbb N}$. This is an {\it exact} alternative to the {\it
approximate} solution ${\widetilde Z}_n^{({\rm sp})}(\epsilon)$.
{\it Equation (\ref{Toda}), known as the Toda lattice equation
\cite{Toda} in the theory of integrable hierarchies \cite{M-1995},
is the first indication of exact solvability hidden in replica
field theories}.

(ii) The second observation borrowed from Ref. \cite{FW-2001}
concerns the fact that, miraculously, the same Toda lattice
equation governs the behaviour of so-called $\tau$-functions
arising in the Hamiltonian formulation of the six Painlev\'e
equations (PI--PVI), which are yet another fundamental object in
the theory of nonlinear integrable systems. Complementary to
(\ref{Toda}), and also luckily, the Painlev\'e equations contain
the hierarchy (or replica) index $n$ as a {\it parameter}. For
this reason, they serve as a proper starting point to build a
consistent analytic continuation of nonperturbative replica
partition functions away from $n$ integers. {\it This Painlev\'e
reduction confirms exact solvability of replica $\sigma$ models
and assists performing the replica limit (\ref{p-cf})}.

{\it GUE density of states revisited}.---With these observations
in hand, let us consider the one-point Green's function for ${\rm
GUE}_N$ where we have a rare luxury of examining the replica
partition function as a function of energy $\epsilon$, replica
parameter $n$, and matrix dimension $N$ both {\it before} and {\it
after} replica $\sigma$ model mapping.

{\it After} replica mapping, the Painlev\'e reduction of the
partition function ${\widetilde Z}_n(\epsilon)$ obeying
(\ref{n-factorisation}) and (\ref{Toda}) with ${\widetilde
Z}_0(\epsilon) \equiv 1$ and ${\widetilde Z}_1(\epsilon) =
H_N(\epsilon)$ [$H_N$ is the Hermite polynomial, see
(\ref{GUE-Z-nfold}) at $n=1$] materialises in the exact
representation \cite{FW-2001}
\begin  {eqnarray}
        \label{P-IV}
        {\widetilde Z}_n(\epsilon) =
        {\widetilde Z}_n(0) \,
        \exp \left(
        \int_0^{i\epsilon} dt\, \varphi_{\rm IV}(t)
        \right), \;\; n \in {\mathbb N}.
\end    {eqnarray}
It involves the Painlev\'e transcendent $\varphi_{\rm IV}(t)=
\varphi_n(N;t)$ satisfying the Painlev\'e IV equation in the
Jimbo-Miwa-Okamoto form \cite{JM-1981}
\begin  {equation}
        \label{toy-eq-1}
        (\varphi_{\rm IV}^{\prime\prime})^2 - 4 (t \varphi_{\rm IV}^\prime
        - \varphi_{\rm IV})^2
        %\nonumber \\
        + 4 \varphi_{\rm IV}^\prime (\varphi_{\rm IV}^\prime - 2n)
        (\varphi_{\rm IV}^\prime + 2 N)
        = 0.
\end    {equation}
The boundary condition is $ \varphi_{\rm IV}(t) \sim (nN/t)
\left(1 + {\cal O}(t^{-1}) \right) $ as  $t \rightarrow +\infty$.
Note that (\ref{toy-eq-1}), and therefore (\ref{P-IV}),  contains
the replica index $n$ as a {\it parameter}.

By derivation, Eq. (\ref{P-IV}) holds for $n$ positive integers
only and, generically, there is no {\it a priori} reason to expect
it to stay valid away from $n \in {\mathbb N}$. We claim, however,
that it {\it is} legitimate to extend (\ref{P-IV}) and
(\ref{toy-eq-1}), as they stand, beyond $n \in {\mathbb N}$ and
consider this extension as a sought analytic continuation. To
prove this, we examine the partition function $Z_n(\epsilon)$ {\it
prior} to $\sigma$ model mapping as given by (\ref{GUE-Z}). In the
eigenvalue representation \cite{M-1991}, Eq. (\ref{GUE-Z}) reduces
to the $N$-fold integral akin to (\ref{GUE-Z-nfold}),
\begin  {eqnarray}
        \label{GUE-eigen}
        Z_n(\epsilon) = \int_{-\infty}^{+\infty} \prod_{k=1}^N
        d\lambda_k e^{-\lambda_k^2} (\lambda_k-\epsilon)^n
        \Delta_N^2(\lambda).
\end    {eqnarray}
Similar to (\ref{GUE-Z-nfold})--(\ref{Toda}), the appearance of
$\Delta_N^2$ in (\ref{GUE-eigen}) leads to the H\"ankel
determinant representation $Z_n(\epsilon) = Z_n(\epsilon;N) = e^{-
N \epsilon^2} \tau_N(\epsilon;n)$ with $\tau_N(\epsilon;n) =
    \det
    \left[
    \partial^{k+\ell}_\epsilon\, \tau_1(\epsilon; n)
    \right]_{k,\ell = 0, \cdots, N-1}$.
Given the initial conditions $\tau_0 \equiv 1$ and
$\tau_1=e^{\epsilon^2} Z_n(\epsilon;1)$, the Toda lattice equation
$\tau_N \, \tau^{\prime\prime}_N
    -
    (\tau^\prime_N)^2 = \tau_{N-1}
    \tau_{N+1}$,
with $N \ge 1$, follows by the Darboux theorem \cite{Darboux}.
Since $n$ is allowed to be an arbitrary complex number in
(\ref{GUE-Z}), the Toda equation determines a whole set of replica
partition functions $Z_n(\epsilon)$ for {\it all} $n \in {\mathbb
C}$. For $Z_n(\epsilon;0) \equiv 1$ and $Z_n(\epsilon;1)=H_n(i
\epsilon)$ (see (\ref{GUE-eigen}) at $N=1$), the Painlev\'e
reduction of the above Toda equation reads \cite{FW-2001}
\begin  {eqnarray}
        \label{before}
        Z_n(\epsilon) = Z_n(0)\, \exp \left(
        \int_0^{\epsilon} dt\, \psi_{\rm IV}(t)
        \right), \;\; n \in {\mathbb C}.
\end    {eqnarray}
Here, $\psi_{\rm IV}(t) = \psi_N(n;t)$ satisfies the Painlev\'e IV
equation
\begin  {equation}
        \label{toy-eq-2}
        (\psi_{\rm IV}^{\prime\prime})^2 - 4 (t \psi_{\rm IV}^\prime
        - \psi_{\rm IV})^2
        %\nonumber \\
        + 4 \psi_{\rm IV}^\prime (\psi_{\rm IV}^\prime - 2N)
        (\psi_{\rm IV}^\prime + 2 n)
        = 0
\end    {equation}
and matches the asymptotic behavior $ \psi_{\rm IV}(t) \sim
    (nN/t) \left(1 + {\cal O}(t^{-1}) \right)
$ at infinity $t \rightarrow + \infty$.

A brief inspection reveals that (\ref{before}) reduces to
(\ref{P-IV}) because of the duality
$
    \left. i \varphi_n(N; it) \right|_{n \in {\mathbb N}} =
    \psi_N(n;t)
$ that can easily be verified from (\ref{toy-eq-1}) and
(\ref{toy-eq-2}) and from the boundary conditions at infinity. As
the above duality between (\ref{toy-eq-1}) and (\ref{toy-eq-2})
formally holds beyond $n \in {\mathbb N}$, we conclude that
(\ref{P-IV}) and (\ref{toy-eq-1}) considered, as they stand, at
arbitrary complex $n$ furnish a proper analytic continuation.
Then, it can be shown \cite{EK-2002} that, in the large-$N$ limit,
the replica projection $G(\epsilon) = \lim_{n\rightarrow 0}
    n^{-1} \partial {\widetilde Z}_n(\epsilon)/\partial \epsilon$
of the so-continued partition function (\ref{P-IV}) results in the
correct expression
$    G(\epsilon)
    = \epsilon - \sqrt{\epsilon^2 - 2N}
$ for the one-point Green's function $G(\epsilon)$. The famous
Wigner's semicircle \cite{M-1991} for the level density
$\nu_N(\epsilon)=-\pi^{-1}{\rm Im} \, G(\epsilon)=
\pi^{-1}\sqrt{2N-\epsilon^2}$ readily follows.

{\it Wigner-Dyson correlations in GUE}.---The two-level
correlation function in GUE (symmetry class \texttt{A} in Cartan
classification) can be treated along the same lines. Defined in
terms of the two-point Green's function $G(s=\epsilon_1 -
\epsilon_2) = G(\epsilon_1,\epsilon_2)$ as $
    R(s) = (1/2) \left[ {\rm Re} \, G(s) -1 \right]
$, it can be obtained from the replica limit $
    G(s) = - \lim_{n\rightarrow 0} n^{-2}
    \partial^2 {\widetilde Z}_n(s)/\partial s^2
$; ${\widetilde Z}_n(s)$ is the fermionic partition function.
Taken at imaginary argument, it is given by the
Verbaarschot-Zirnbauer integral \cite{VZ-1985}
\begin  {eqnarray}
        \label{vz}
        {\widetilde Z}_n(is) = \frac{e^{ns}}{s^{n^2}}
        \int_{0}^{2s} \prod_{k=1}^n d\lambda_k
        e^{- \lambda_k} \Delta_n^2(\lambda).
\end    {eqnarray}
This is a Fredholm determinant \cite{TW-1994} associated with a
gap formation probability \cite{M-1991} within the interval
$(2s,+\infty)$ in the spectrum of an auxiliary $n \times n$
Laguerre unitary ensemble. Utilising the results of Refs.
\cite{TW-1994,FW-2002} we derive \cite{EK-2002}
\begin  {eqnarray}
        \label{pV-exp}
        {\widetilde Z}_n(is) = \exp
        \left(
        \int_0^{2s} dt \frac{\sigma_{\rm V}(t) - n^2 + nt/2}{t}
        \right).
\end    {eqnarray}
Here, $\sigma_{\rm V}(t) = \sigma(n;t)$ satisfies the
Jimbo-Miwa-Okamoto form of Painlev\'e V equation \cite{JM-1980}
\begin  {equation}
        \label{pV}
        (t \sigma^{\prime\prime}_{\rm V})^2 -  (\sigma_{\rm V} - t
        \sigma^\prime_{\rm V})
        %\nonumber \\
        %&\times&
        [\sigma_{\rm V} - t \sigma^\prime_{\rm V} + 4 \sigma^\prime_{\rm V}
        (\sigma^\prime_{\rm V}+ n)]
        =0
\end    {equation}
with the boundary condition \cite{B-2002,EK-2002} $ \sigma_{\rm
V}(t) \sim n^2 e^{-t}/t$ as $t \rightarrow + \infty$.

The replica limit is governed by the behaviour of ${\widetilde
Z}_n(s)$ in the vicinity of $n=0$. In concert with the above
discussion, we assume that (\ref{pV-exp}) and (\ref{pV}) determine
the desired analytic continuation. Expanding the solution to
(\ref{pV}) around $n=0$ yields \cite{EK-2002} $
        \label{s5-expan}
        \sigma_{\rm V}(t) = n^2 E_2(t) + {\cal O}(n^3)$,
where $E_2(t)$ is the exponential integral
$
 E_\ell(z) = \int_1^\infty dt \, e^{-zt}\, t^{-\ell}
$, $\, {\rm Re}\, z >0$. As a result, the small-$n$ expansion of
the partition function reads \cite{EK-2002}
\begin  {eqnarray}
        \label{gue-zn-expan}
        \ln {\widetilde Z}_n(is) &=& n s - n^2 \left[
        1 + \gamma + \ln (2s)
        \right. \nonumber\\
        &+&  \left. E_1(2s) - E_2(2s)
        \right]+ {\cal O}(n^3).
\end    {eqnarray}
Here, $\gamma=0.577\dots$ is the Euler constant.

To the best of our knowledge, this is the first nonperturbative
evaluation of ${\widetilde Z}_n$. Implementing the replica limit,
one derives the two-point Green's function of the form $G(s) = 1 +
2i s^{-2}\, \sin(s) e^{is}$. This, in turn, reproduces the
celebrated Wigner-Dyson two-point correlation function
\cite{M-1991} $R(s)= \pi \delta(s) - s^{-2} \sin^2(s)$. Note that
this result holds for arbitrary $s$ down to zero (compare to Ref.
\cite{KM-1999}).

{\it Further examples}.---The same strategy can be applied to
demonstrate exact solvability of fermionic replica $\sigma$ models
for other random matrix ensembles associated with the Toda lattice
hierarchy. Our partial list includes chiral GUE \cite{VZ-1993}
(chGUE, symmetry class \texttt{AIII}) and Ginibre's ensemble
\cite{G-1965} of complex matrices with no further symmetries. A
detailed account of the Painlev\'e reduction for these ensembles
will be presented elsewhere \cite{EK-2002}. Here we announce only
small-$n$ expansions of the corresponding nonperturbative replica
partition functions. Adopting the notation of Ref. \cite{DV-2001}
for chGUE [their Eq. (12)] and of Ref. \cite{NK-2002} for
Ginibre's matrix model [their Eqs. (13) and (32)], we have derived
\cite{EK-2002} for the chGUE
\begin  {eqnarray}
        \label{chiral-exp}
        \ln {\widetilde Z}_\nu^{(n)}(\epsilon) &=&  n \Big(
        \nu \ln \epsilon +
        \int_0^\epsilon dt \, t \big[
        K_\nu(t)\, I_\nu(t) \nonumber \\
        &+&
        K_{\nu-1}(t) \, I_{\nu+1}(t)
        \big]
        \Big) + {\cal O}(n^2)
\end    {eqnarray}
by the Painlev\'e III reduction ($I_\nu$ and $K_\nu$ are modified
Bessel functions) whilst for Ginibre's complex matrices
\cite{Remark5}
\begin  {widetext}
\begin  {eqnarray}
        \label{gin-exp}
        \ln {\widetilde Z}_n(z,{\bar z}) &=&
        n(z{\bar z}) \left[
        1 + \frac{(z{\bar z})^N}{(N+1)! (N+1)} \,
        {}_2 F_2(N+1,N; N+2,N+2; -z{\bar z})
        \right] + {\cal O}(n^2),\\
        \label{gin-exp-2}
        \ln {\widetilde Z}_n(z,{\bar z}; \omega, {\bar \omega}) &=&
        n \left( 2 z{\bar z} + \frac{\omega {\bar \omega}}{2} \right)
        - n^2 \left[
        1 + \gamma + \ln (\omega {\bar \omega})
        +  E_1(\omega{\bar \omega}) - E_2(\omega{\bar \omega})
        \right]+ {\cal O}(n^3)
\end    {eqnarray}
\end    {widetext}
by the Painlev\'e V reduction (${}_2 F_2$ is a hypergeometric
function). Exact correlation functions for the above ensembles
follow from (\ref{chiral-exp}) -- (\ref{gin-exp-2}) upon
implementing proper replica limits. Out of three, the small-$n$
expansion (\ref{chiral-exp}) is of particular interest as it holds
for arbitrary values of the topological charge $\nu$. Alternative
calculational scheme \cite{DV-2001} based on the saddle point
evaluation of ${\rm U}(n)$ matrix integrals, much in line with
\cite{KM-1999}, could reproduce exact results for $\nu$ half
integers only, with exactness being secured\cite{Z-1999} by the
Duistermaat-Heckman theorem\cite{DH-1982}.

(i) The \texttt{A}, \texttt{AIII} and Ginibre's random matrix
models exhaust our list of ensembles illustrating exact
solvability of fermionic replica field theories. The integrable
structure of all of them (as well as of those for the matrix
models from {\texttt B}, {\texttt C}, and {\texttt D} Cartan
symmetry classes not considered in the Letter) is related to the
Toda lattice (\ref{Toda}) whose appearance is traced back to a
particular $\beta=2$ symmetry of the corresponding replica
partition functions.

(ii) We expect the other random matrix ensembles exhibiting
$\beta=1$ and $\beta=4$ symmetries (belonging to \texttt{AI},
\texttt{BDI}, \texttt{CI} and \texttt{AII}, \texttt{CII},
\texttt{DIII} classes in Cartan classification, respectively) to
be exactly solvable as well. In those cases, integrable
hierarchies related to the Pfaff lattice \cite{AM-2001} are likely
to arise.

(iii) Another application of the formalism developed would be
getting further insight into controversies surrounding bosonic
replicas which are known to be a total failure in the description
\cite{VZ-1985} of spectral correlations in some ensembles while
being quite successful in the description of others
\cite{NK-2002}.

(iv) Finally, it would be desirable to figure out to what extent
the nonperturbative Painlev\'e reduction of replica partition
functions reported in this Letter is helpful in the replica
treatment \cite{F-1983} of recently advocated universal
``zero-dimensional'' random Hamiltonians which include
interactions \cite{Ints}.

I appreciate a clarifying correspondence with Peter J. Forrester.
Craig A. Tracy is thanked for helpful conversations during a
workshop on random matrices in SUNY at Stony Brook in February
2002.


\begin{references}
\bibitem{Remark1} Throughout this paper, no care is taken of
all sorts of overall constants which tend to unity in the replica
limit.
\end{references}
\end{document}